\documentclass[preprint]{aastex}
\usepackage{epsfig}
\usepackage{subfigure}

\newbox\grsign \setbox\grsign=\hbox{$>$} \newdimen\grdimen \grdimen=\ht\grsign
\newbox\simlessbox \newbox\simgreatbox
\setbox\simgreatbox=\hbox{\raise.5ex\hbox{$>$}\llap
     {\lower.5ex\hbox{$\sim$}}}\ht1=\grdimen\dp1=0pt
\setbox\simlessbox=\hbox{\raise.5ex\hbox{$<$}\llap
     {\lower.5ex\hbox{$\sim$}}}\ht2=\grdimen\dp2=0pt

\def\simless{\mathrel{\copy\simlessbox}}

\shorttitle{Predictions of mixed density fields in CMBR}
\shortauthors{Andrade et al.}

\begin{document}

\title{Predictions of mixed non-Gaussian cosmological density
fields for the cosmic microwave background radiation}

\author{A.P.A. Andrade, C.A. Wuensche}
\affil{Divis\~ao de Astrof\'{\i}sica, Instituto Nacional de
Pesquisas Espaciais/MCT} \affil{CP 515, 12201-970 SP, Brazil}
\email{apaula@das.inpe.br, alex@das.inpe.br}\and
\author{ A.L.B. Ribeiro} \affil{Departamento de
Ci\^encias Exatas e Tecnol\'ogicas, Universidade Estadual de Santa
Cruz} \affil{CP 45650-000 BA, Brazil}\email{albr@uesc.br}

\begin{abstract}
We present simulations of the Cosmic Microwave Background
Radiation (CMBR) power spectrum for a class of mixed,
non-Gaussian, primordial random fields. We assume a skew positive
mixed model with adiabatic inflation perturbations plus additional
isocurvature perturbations possibly produced by topological
defects. The joint probability distribution used in this context
is a weighted combination of Gaussian and non-Gaussian random
fields, such as: P(\ensuremath{\delta}) =
(1-\ensuremath{\alpha})f$_{1}$(\ensuremath{\delta}) +
\ensuremath{\alpha}f$_{2}$(\ensuremath{\delta}), where
f$_{1}$(\ensuremath{\delta}) is a Gaussian distribution,
f$_{2}$(\ensuremath{\delta}) is a non-Gaussian general
distribution, and \ensuremath{\alpha} is a scale dependent mixture
parameter. Results from simulations of CMBR temperature and
polarisation power spectra show a distinct signature for very
small deviations ($\simless$ 0.1\%) from a pure Gaussian field. We
discuss the main properties of such mixed models as well, their
predictions and suggestions on how to apply them to small scale
CMBR observations. A reduced $\chi^2$ test shows that the
contribution of an isocurvature fluctuation field is not ruled out
in actual CMB observations, even in WMAP first-year sky map.

\end{abstract}


\section{Introduction}
One of the main goals of cosmology today is to determine the
origin of primordial density fluctuations. Since CMBR carries the
intrinsic statistical properties of cosmological perturbations, it
is considered the most powerful tool to investigate the nature of
cosmic structure. Tests for Gaussianity of CMBR anisotropy can
discriminate between various cosmological models for structure
formation.

The most accepted model for structure formation assumes initial
quantum fluctuations created during inflation and amplified by
gravitational effects. The standard inflation model predicts an
adiabatic uncorrelated random field with a nearly flat,
scale-invariant spectrum on scales larger than \ensuremath{\sim}
1-2\ensuremath{^\circ} (Guth 1981; Salopek, Bond \& Bardeen 1981;
Bardeen, Steinhardt \& Turner 1983).
Simple inflationary models also predict the random field
follows a nearly-Gaussian distribution, where just small
deviations from Gaussianity are allowed (e.g. Gangui et al. 1994).
However, larger deviations are also possible in a wide class of
alternative models like non standard inflation models with a
massless axion field (Allen, Grinstein \& Wise 1987); with
multiple scalar fields (Salopek, Bond \& Bardeen 1981); with a
massive scalar field (Koyama, Soda \& Taruya 1999); with
variations in the Hubble parameter (Barrow \& Coles 1990).
Cosmic defects models (Kibble 1976; Magueijo \& Brandenbergher
2000) also predict the creation of non-Gaussian random fields. In
hybrid inflation models (Battye \& Weller 1998; Battye, Magueijo
\& Weller 1999), structure is formed by a combination of (inflation-produced) adiabatic and (topological defects induced) isocurvature density
fluctuations. The topological defects are assumed  to appear during the phase
transition that marks the end of the inflationary epoch. In this
scenario, the fields are uncorrelated, combined by a weighted
average and obey a non-Gaussian statistics.

The interest in non-Gaussian structure formation models is not
unjustified. Indeed, the increasing  number of galaxies observed
at high redshifts clearly disfavors standard inflationary  models
with Gaussian initial conditions, which predict these objects
should be very rare in the universe (e.g. Weymann et. al. 1998).
In addition, statistical evidence for a small level of
non-Gaussianity in the anisotropy of CMBR has been found in the
COBE/DMR four-year sky maps (Ferreira, Magueijo \& G\'{o}rski
1998) and in the WMAP first-year observations (Chiang et al.
2003). On the other hand, recent CMBR anisotropy observations in
large (Smoot et al. 1992; Bennett et al. 1996), intermediate (de
Bernardis et al. 2000; Hanany et al. 2000) and small angular
scales (Halverson et al. 2002; Stompor et al. 2001; Mason et al.
2002; Pearson et al. 2002; Peiris et al. 2003; Kogut et al. 2003)
seem to be in reasonable agreement with inflation predictions.
Hence, it is hard to either accept or rule out the non-Gaussian
contribution to structure formation.

Actually, if one looks at all the available data, one possible
interpretation suggests an intermediate situation where realistic
initial conditions for structure formation have small but
significant departures from Gaussianity. A scenario in which the
details of such departures are understood and calculated may offer
a new alternative for evolution of cosmological perturbations (for
a present overview, see e.g. Gordon 2001; Gordon et al. 2001) and
formation of structures in the Universe. Hopefully, with the
improved quality of CMBR observations both on the ground and on
board stratospheric balloons, and with data coming from MAP and
PLANCK satellites missions (Tauber 2000; Wright 2000) in a very
near future, we will possibly be able to unveil the statistical
properties of the density field with good precision. This will
finally allow us to choose, among the large number of available
candidates, the cosmological models that adequately fit the
observational data.

In this work, we explore the hypothesis that the initial
conditions for structure formation do not necessarily build  a
single, one-component random field, but a weighted combination of
two or more fields. In particular, we are interested in simple
mixtures of two fields, one of them being a dominant Gaussian
process. In a previous work, Ribeiro, Wuensche \& Letelier (2000)
(hereafter RWL) used such a model to probe the galaxy cluster
abundance evolution in the universe and found that even a very
small level of non-Gaussianity in the mixed field may introduce
significant changes in the cluster abundance rate. Now, we
investigate the effects of mixed models in the CMBR power
spectrum, considering a general class of finite mixtures and
always combining a Gaussian with a second field to produce a
positive skewness density fluctuation field. For this combination
we adopt a scale dependent mixture parameter and a power-law
initial spectrum, $P(k)=Ak^n$. CMBR temperature and polarisation
power spectra are simulated for a flat \ensuremath{\Lambda}-CDM
model, while varying some cosmological parameters, and the
temperature fluctuations are estimated. We show how the shape and
amplitude of the fluctuations in CMBR are dependent on such mixed
fields and how we can distinguish a standard adiabatic Gaussian
field from a mixed non-Gaussian field.

This paper is organised as follows: in the next section, we
discuss the main properties of the mixed models. In section 3, we
present the simulations results for a standard cosmological
\ensuremath{\Lambda}-CDM model mixing Gaussian, lognormal,
exponential, Maxwellian and Chi-squared distributions. Simulations
for different combinations of cosmological parameters are
presented in section 4. In section 5, we finally summarise and
discuss the possibilities of use of the proposed mixed composition
model for parameter estimation in future small scale CMBR
observations.

\section{General Description}

\subsection{Non-Gaussian Random Fields}

The use of statistical methods to describe the structure formation
process in the universe is due to the lack of complete knowledge
about the density fluctuation field $\delta({\bf x})$ at any time
$t$. This lead us to treat $\delta({\bf x})$ as a random field in
the 3D-space and to assume the universe as a random realization
from a statistical ensemble of possible universes. In general, it
is possible to assure Gaussianity to this field by simply invoking
the central limit theorem (CLT). However, in order to better
understand the process of structure formation, it is necessary to
investigate the existence and (in case it exists) contribution of non-Gaussian effects to the
primordial density field.

Non-gaussianity implies an infinite range of possible statistical
models. Hence, the usual approach to this subject is to examine
specific classes of non-Gaussian fields. The general procedure to
create a wide class of non-Gaussian models is to admit the
existence of an operator which transforms Gaussianity into
non-Gaussianity according to a specific rule. For a small level of
non-Gaussianity, the perturbation theory works well for the
density field. For instance, we can define a zero-mean random
field $\psi$ which follows a local transformation $\mathcal{F}$ on
a underlying Gaussian field:

\begin{equation}
\psi({\bf x}) = \mathcal{F}[\phi]\equiv \alpha\phi({\bf x}) + \epsilon
{(\phi^{2}({\bf x}) - \langle\phi^{2}({\bf x})\rangle)}
\end{equation}

\noindent where $\alpha$ and $\epsilon$ are free parameters of the
model. In the limit $\alpha\rightarrow 0$, $\psi$ is chi-squared
distributed, while for $\epsilon\rightarrow 0$ one recovers the
Gaussian field. The field described by $\psi$ is physically
motivated in the context of non-standard inflation models (e.g.
Falk et al. 1993; Gangui et al. 1994). Besides, transformation (1)
may be considered as a Taylor expansion of more general
non-Gaussian fields (e.g. Coles \& Barrow 1987; Verde et al.
2000). We also should note that $P(\phi)$ is a Gaussian PDF, while
$P(\psi)=\int W(\psi|\phi)P(\phi)\;d\phi$, where $W(\psi|\phi)$ is
the transition probability from $\phi$ to $\psi$ (e.g. Taylor \&
Watts 2000; Matarrese, Verde \& Jimenez 2000).

An alternative approach to study non-Gaussian fields is that
proposed by RWL, in which the PDF itself is modified as a mixture:
$P(\psi)= \alpha f_1(\phi) + (1-\alpha)f_2(\phi)$, where
$f_1(\phi)$ is a (dominant) Gaussian PDF and $f_2(\phi)$ is a
second distribution, with $\alpha$ being a parameter between 0 and
1. This parameter gives the absolute level of Gaussian deviation,
while $f_2(\phi)$ modulates the shape of the resultant
non-Gaussian distribution. The particular choice of RWL was to
define $f_2$ as a lognormal distribution. Instead of supporting
this model with a specific inflation picture, the authors take the
simple argument that the real PDF of the density field cannot be
strongly non-Gaussian (from COBE data constraints, see Stompor et
al. 2001, and from WMAP, see Komatsu et al. 2003) and, at the same
time, it should be approximately log-normal in the non-linear
regime (from Abell/ACO clusters data, see Plionis \& Valdarnini
1995). Indeed, Coles \& Jones (1991) argued that the lognormal
distribution provides a natural description for the density
fluctuation field resulting from Gaussian initial conditions in
the weakly non-linear regime. Hence, RWL envisage $\alpha$ as a
function of time which turns a nearly Gaussian PDF at
recombination ($\alpha\approx 1$) into a nearly lognormal
distribution ($\alpha\approx 0$) over the non-linear regime. The
problem of taking a direct relationship between the PDF at two
different times is not considered by RWL, but it could be done in
the context of the extended perturbation theory (Colombi et al.
1997). It is important to note that a mixture of distributions
including a lognormal component implies the existence of a
non-perturbative contribution of type $e^\phi$ in the primordial
density field, such that the transformation $\mathcal{F}$ becomes:

\begin{equation}
\psi ({\bf x}) = \mathcal{F}[\phi] = \alpha\phi({\bf x})
+ (1-\alpha )e^{\phi({\bf x})}
\end{equation}

\noindent The physics of the field described by $\psi$ is studied
elsewhere (Ribeiro et al. 2003). Here, in continuity of the work
of RWL, we investigate the implications of mixed models for the
CMBR power spectrum. We take the attitude that, in face of the
difficulties to completely describe the primordial density field
and its evolution, it is valid to take the predictions of a
tentative model like RWL and compare them with observations. Our
primary aim is just to find a successful and simple idealization
for observed phenomena. Next, we describe the technical details of
the mixed models and how to use them to make predictions for the
CMBR.

\subsection{The Mixed Models}

A Gaussian random field is one in which the Fourier components
$\delta_{k}$ have independent, random and uniformly distributed
phases. In this case, the probability density function in Fourier
Space is:

\begin{equation}
P[\delta_k]~\propto ~{\rm exp}~\left(-{1\over
2}\sum_k{|\delta_k|^2\over \sigma_k^2}\right)
\end{equation}

Such a condition means that the phases are non-correlated in space
and assures the statistical properties of the Gaussian fields are
completely specified by the two-point correlation function or,
equivalently, by its power spectrum $P(k)=|\delta_k|^{2}$, which
contain information about the density fluctuation amplitude of
each scale k. In an isotropic and homogeneous Universe, $k$
represents only the wavevector amplitude.

In the mixed scenario, we suppose that the field has a probability
density function of the form:

\begin{equation}
P[\delta_k]~\propto ~(1-\alpha)f_1(\delta_k) + \alpha
f_2(\delta_k)
\end{equation}

\noindent In general, the PDF of the Fourier components
P$[\delta_k]$ is not equal to the PDF of the field P$[\delta]$.
However, we can always consider a non-gaussian distribution field
like a combination of a gaussian and a non-gaussian distribution.
This means that a mixed distribution can be applied to even real
and Fourier space in a non-gaussian context, but this not mean
that the real and Fourier space have the same kind of combination.
In the mixed context, the Fourier components $\delta_{k}$ have
only a small fraction of correlated phase in space represented by
the second distribution. The first field will always be the
Gaussian component and a possible effect of the second component
is to modify the Gaussian field to have a positive tail. The
parameter $\alpha$ in (4) allows us to modulate the contribution
of each component to the resultant field. The Gaussian component
represents the adiabatic (or isentropic) inflation field and the
second component may represent the effect of adding an
isocurvature field produced due to some primordial mechanism
acting on the energy distribution, such as topological defects.
The two component random field can be generated by taking
$\delta_{k}= P(k)\nu^2$, where $\nu$ is a random number with
distribution given by (4). Then, the mean fluctuation,
$\langle\delta^{2} (\ensuremath{x})\rangle$, is proportional to:

\begin{equation}
\int_k P(k)\left[\int_\nu[(1-\alpha)f_1(\nu) + \alpha
f_2(\nu)]\nu^2\;d\nu\right]\;d^3k
\end{equation}

\noindent The primordial power spectrum of the mixed field has the
form:

\begin{equation}
P(k)^{mix} \equiv M^{mix} (\alpha)P(k)
\end{equation}

\noindent where the $P(k)$ represents a power-law spectrum and
M$^{mix}$($\alpha$) is the mixture term, which account for the
statistics effect of a new component, a functional of
$f_{\mathit{1}}$ and $f_{\mathit{2}}$:

\begin{equation}
M^{mix}(\alpha)\equiv\int_\nu[(1-\alpha)f_1(\nu)+\alpha
f_2(\nu)]\nu^2\; d\nu
\end{equation}

\noindent Resolving this integral assuming $f_{1}$ as the Gaussian
distribution, the mixture term is:

\begin{equation}
M^{mix}(\alpha)=1-\alpha + \alpha\int_\nu f_2(\nu)\nu^2\; d\nu
\end{equation}

In this work, we explore the case of a positive skewness model,
where the second field adds to the Gaussian component a positive
tail representing a number of rare peaks in the density
fluctuation field. To represent the effect of adding an
isocurvature field we have chosen the well-known lognormal,
exponential, Rayleigh, Maxwellian and Chi-squared distributions as
the second component. These random fields have already been used
to calculate the size and number of positive and negative peaks in
CMBR distribution, under the assumption that it possesses a
single, non-Gaussian, component (Coles \& Barrow 1987), but, to
our present knowledge, they have never been used in this context
of mixed fields.

Like the hybrid inflation models (Battye \& Weller 1998; Battye,
Magueijo \& Weller 1999), mixture models consider the scenario in
which structure is formed by both adiabatic density fluctuations
produced during inflation and active isocurvature perturbations
created by cosmic defects during a phase transition which marks
the end of inflationary epoch. Nevertheless, the mixed scenario
considers a possible correlation between the adiabatic and the
isocurvature fields only on the post-inflation Universe. So, the
fluctuations in super-horizon scales are strictly uncorrelated.
While the hybrid inflation scenario considers the super and
sub-degree scales of CMBR anisotropy due to uncorrelated strings
and inflation fields, the mixed model considers an effective mixed
correlated field acting inside the Hubble horizon, in sub-degree
scales. To allow for this condition and keep a continuous mixed
field, the mixture parameter was defined as a scale dependent
parameter, \ensuremath{\alpha} \ensuremath{\equiv}
\ensuremath{\alpha}(k). The simplest choice of
\ensuremath{\alpha}(k) is a linear function of $k$:

\begin{equation}
\alpha(k) \equiv \alpha_{0}k
\end{equation}

\noindent In this case, the mixture term is a function of
$k$, $M(\alpha_{0}$, $k$), and the mixed primordial power spectrum
is:

\begin{eqnarray}
P(k)^{mix}\equiv M^{mix}(\alpha_{0},\textit{k})P(k) \nonumber\\
= k^{n} + M(\alpha_{0})k^{n+1}
\end{eqnarray}

\noindent where M(\ensuremath{\alpha}$_{0}$) represents only the
coefficient dependence, \ensuremath{\alpha}$_{0}$. In the case of
a pure Gaussian field, \ensuremath{\alpha}$_{0}$
\ensuremath{\approx} 0, and the mixed power spectrum will be a
simple power-law spectrum. In the case of a mixed field, the phase
correlations between both fields are estimated by the integral in
Eq. (8), on mixture scales defined by Eq. (9).

\section{Mixed Non-Gaussian Fields in the Cosmic Microwave Background Radiation}
Since radiation and matter were tightly coupled up to $3\times
10^5$ years, understanding the statistical properties of the CMBR
can be an extremely powerful tool to investigate the Gaussian
nature of the cosmological density fluctuations field. Comparing
theoretical predictions and observational data, it is possible to
select the models that account for the best description of the
temperature field. The statistical nature of the initial
conditions is a basic assumption of an algorithm to generate CMBR
power spectra. In this work, instead of assuming the usual
Gaussian initial condition, we analyze the consequences of using
mixed (non-Gaussian) initial conditions to generate CMBR power
spectra.

To estimate a CMB temperature power spectrum, we need to evaluate
the evolution of fluctuations generated in the early universe
through the radiation-dominated era and recombination. In a mixed
model with a small deviation from Gaussianity (i.e. possible small
fraction of correlated phases in space), we can, for simplicity,
ignore the coupled phase evolution through the radiation-dominated
era and consider only the evolution of the amplitude of the
Fourier modes within the framework of linearized perturbation
theory. Following this approach, the evolution of both adiabatic
and isocurvature components of the mixed density field is
considered as an independent process. Only their effective
amplitude correlation is considered at the Last Scattering Surface
(LSS) of the photons from CMBR. To compute the independent
evolution, we have used the Linger function of the COSMICS code
package (Bertschinger 1999) for an adiabatic
\ensuremath{\Lambda}-CDM mode and an isocurvature
\ensuremath{\Lambda}-CDM mode. The Linger function integrates the
coupled, linearized, Einstein, Boltzmann, and fluid equations
governing the evolution of scalar metric perturbations and density
perturbations for photons, baryons and cold dark matter in a
perturbed flat Robertson-Walker universe, in a synchronous gauge.
Linger does generate the photon density field to compute the CMBR
anisotropy.

To allow for a possible coupling of both fields in the last
scattering surface, the CMBR temperature and polarization power
spectra were estimated by a mixed photon density function
incorporated to the original COSMICS package in estimation of the
multipole moments, C$_{l}$:

\begin{equation}
C_l = 4\pi \int_0^{k_{max}}\;d^3k P^{mix}(k)
(\Delta_l^{mix})^2(k,\tau)
\end{equation}

\noindent The function $\Delta_l^{mix}$ represents the mixed
photon density field in the last scattering surface defined by:

\begin{equation}
\Delta_l^{mix}\equiv(1-\alpha_0k)\Delta_l^{Adi} +
\alpha_0k\Delta_l^{Iso}
\end{equation}

\noindent where $\Delta_l^{mix}$ and $\Delta_l^{Iso}$ are the
photon density function estimated by COSMICS, respectively, for an
adiabatic and an isocurvature seed initial conditions.

Since we are not considering the correlation between modes in
different scales (while we are considering just linear evolution)
the amplitude of the mixed field is obtained by the moment:

\begin{eqnarray}
\langle\Delta_l^{mix}(k)\Delta_l^{mix}(k)^{*}\rangle=\langle |\Delta_l^{mix}|^2\rangle= \langle|\Delta_l^{Adi}|^2 +\nonumber\\
\ \alpha_0^2k^2( |\Delta_l^{Iso}|^2 +
|\Delta_l^{Adi}|^2 -2|\Delta_l^{Iso}||\Delta_l^{Adi}| ) + \nonumber\\
2\alpha_0k|\Delta_l^{Adi}||\Delta_l^{Iso}|
-2\alpha_0k|\Delta_l^{Adi}|^2\rangle \nonumber\\
\end{eqnarray}

\noindent which contains correlated terms between the amplitudes
of the adiabatic and the isocurvature fields, and not only the
independent contribution of each field. The coefficients of the
amplitudes appearing on the second and third line of Eq. (13)
represent the mixing effect between the fields and the mixing
parameter \ensuremath{\alpha}$_{0}$ controls the contribution of
the isocurvature field relative to the adiabatic field. Since we
are not considering independent fields amplitudes and we also
consider a possible small fraction of phase correlation, we can
describe this mixed field by a non-gaussian statistics like Eq.
(4).

Inserting (13) in (11), we have a mixed term in the C$_{l}$
estimation. This condition suggests that the amplitude of both
fields are cross-correlated at the LSS, with a mixing ratio
defined by \ensuremath{\alpha}$_{0}$ and in a characteristic scale
defined by \ensuremath{\alpha}$_{0}$\textit{k}. The power spectra
estimated by Eq. (11) consider a flat \ensuremath{\Lambda}-CDM
universe distorted by a non-gaussian statistics, with constant
spectral index in the range of 0.8 \ensuremath{\leq} n
\ensuremath{\leq}1.2. In a case of a pure gaussian field,
\ensuremath{\alpha}$_{0}$=0, we obtain only the first term of the
power law spectrum in Eq. (10), for a pure and non-correlated
adiabatic field, the first term in Eq. (13), described only by a
gaussian distribution.

In this section we present the simulations of the CMBR temperature
and polarization power spectra for the present standard
\ensuremath{\Lambda}-CDM model (H$_{0}$= 70Km/sMpc;
\ensuremath{\Omega}$_{0}$ = 1; \ensuremath{\Omega}$_{b}$h$^{2}$ =
0.03, \ensuremath{\Omega}$_{CDM}$h$^{2}$ = 0.27,
\ensuremath{\Omega}$_{\ensuremath{\Lambda}}$h$^{2}$ = 0.7 and n
=1.1). Another set of simulations, using different values for the
above parameters, is presented in Section 4. Figure 1 contains the
mixed (Gaussian-lognormal) CMBR temperature power spectrum
estimated for different values of \ensuremath{\alpha}$_{0}$. The
temperature fluctuations are normalized by the RMS quadrupole
amplitude estimated by the COBE/DMR experiment, Q$_{rms}$=( 13
\ensuremath{\pm} 4) \ensuremath{\mu}K (Smoot et al. 1992). Figure
2 shows the mixed (Gaussian-lognormal) polarization power
spectrum.

\begin{figure}
\begin{center}
\plotone{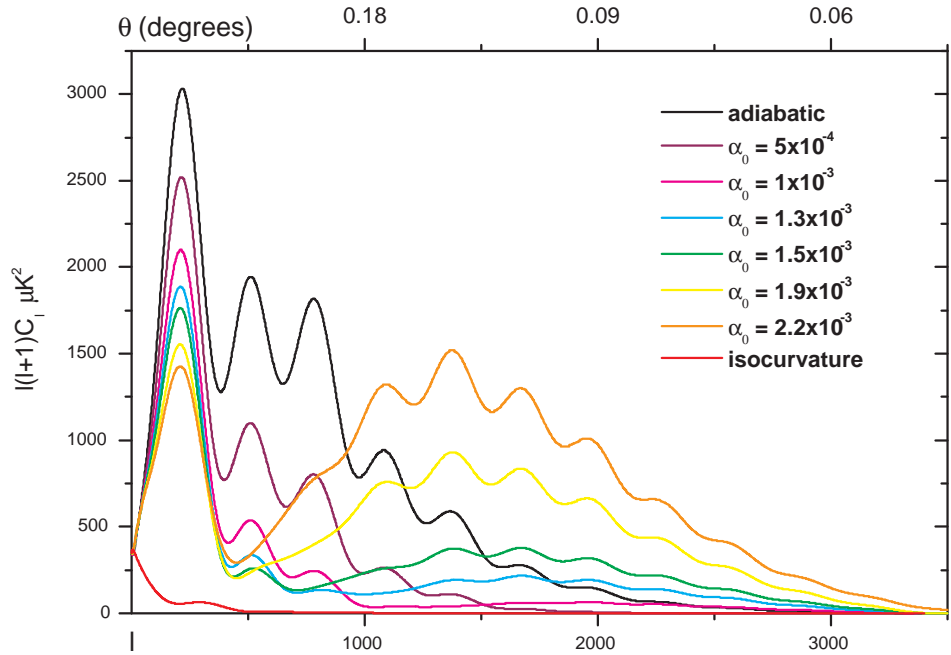} \caption{\small CMBR temperature mixed angular
power spectrum estimated for a standard \ensuremath{\Lambda}-CDM
model in different mixing degrees. }
\end{center}
\end{figure}

\begin{figure}
\begin{center}
\plotone{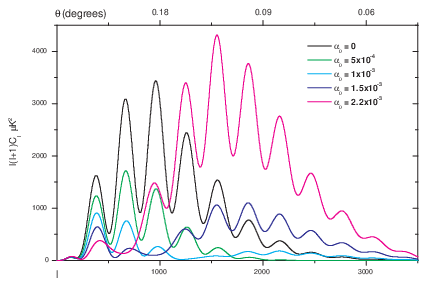} \caption{\small CMBR polarization mixed angular
power spectrum estimated for a standard \ensuremath{\Lambda}-CDM
model in different mixing degrees. }
\end{center}
\end{figure}

In Figures 1 and 2, we see how the shape and amplitude of the
spectrum change even for small values of \ensuremath{\alpha}$_{o}$
(\ensuremath{\sim}10$^{-4}$-10$^{-3}$). The peaks intensity are
clearly susceptible to the existence of mixed fields, although
distinguishing peaks of higher order (2$^{nd}$, 3$^{rd}$, etc.) in
the mixed context is not a straightforward task, since their
intensity, compared to the first peak, are very low. But, on the
other hand, it is very clear how the peaks intensity changes while
the mixture increases. The effect of increasing
\ensuremath{\alpha}$_{0}$ is a power transfer to smaller scales (l
\texttt{>}1000), while the super-degree scales are not affected.
According to our definition, the effect of \ensuremath{\alpha}(k)
upon the fields can be seen only inside the Hubble horizon. In
Figure 3, we see that the fraction of the CMBR polarized component
depends on the mixture parameter and ranges from 4.3 to 7.7\% of
the total intensity for 10$^{-4}$\ensuremath{\leq}
\ensuremath{\alpha}$_{0}$ \ensuremath{\leq}10$^{-3}$.

We consider the relation between the peaks' intensity and
\ensuremath{\alpha}$_{0}$ a key result of this work, since it is a
prediction of the mixed model that can be easily tested and,
moreover, offers a good and straightforward tool to be applied to
the forthcoming data sets from present CMBR experiments and WMAP
and PLANCK satellites. Both of them will map the CMBR power
spectrum in the $l-$interval studied in this paper and the data
from the above mentioned experiments will surely offer a good
bench test for this model.

To illustrate the mean properties of the CMBR mixed fluctuations
field, we have estimated the mean temperature fluctuations,
(\ensuremath{\Delta}T/T)$_{rms}$:

\begin{equation}
\left({\Delta T \over T}\right)_{rms}^2 = {1\over
4\pi}\sum_{l=2}{(2l+1)C_l}
\end{equation}

\noindent The behaviour of (\ensuremath{\Delta}T/T)$_{rms}$ for
the Gaussian-lognormal mixture is shown in Figure 4. For large
values of \ensuremath{\alpha}$_{0}$
(\ensuremath{\alpha}$_{0}$\texttt{>}3x10$^{-3}$) we see a fast
increase in the temperature fluctuations, probably caused by
correlation excess between the mixed fields, resulting in more
power in small scales. From these results, we can set an acceptable
range for $\alpha_0$ to be $\alpha_0 \lesssim 3\times 10^{-3}$.

\begin{figure}
\begin{center}
\plotone{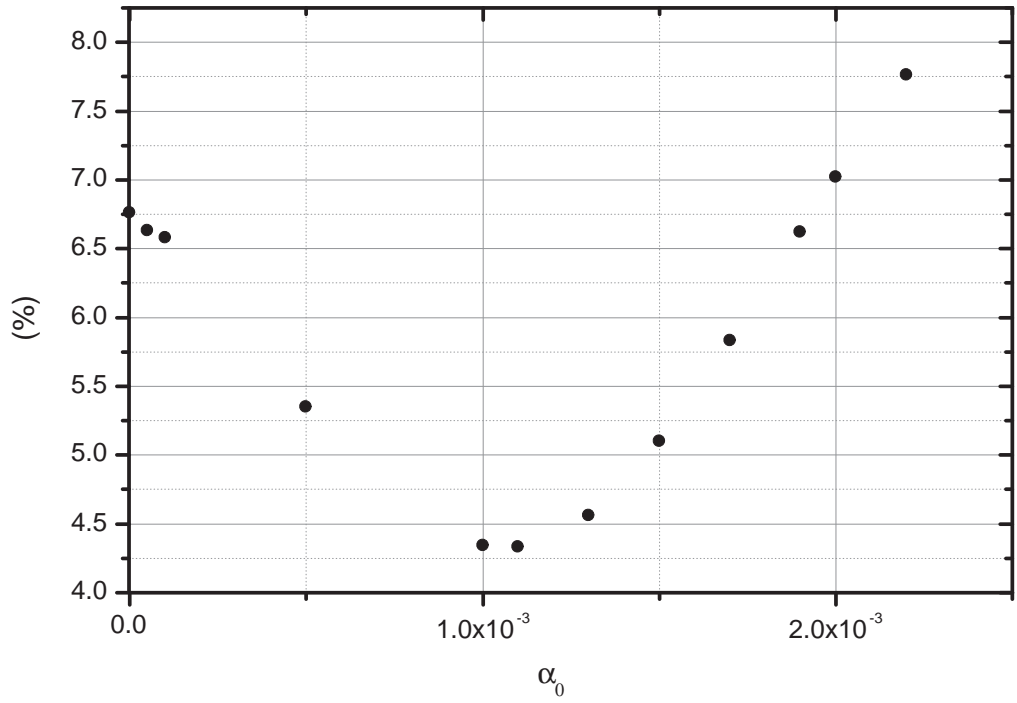} \caption{\small CMBR polarized component
simulated for a mixed Gaussian-lognormal model in a standard
combination of cosmological parameters.}
\end{center}
\end{figure}

\begin{figure}
\begin{center}
\plotone{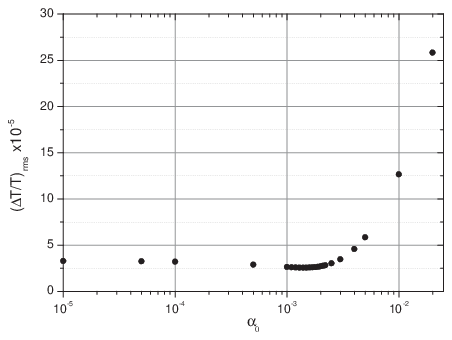} \caption{\small CMBR mean fluctuation temperature
estimated for a Mixed Gaussian-lognormal model in a standard
combination of cosmological parameters.}
\end{center}
\end{figure}

In Table 1, we present the mean temperature fluctuations
(\ensuremath{\Delta}T/T)$_{rms}$ estimated for different values of
\ensuremath{\alpha}$_{0}$ for mixtures between Gaussian,
Exponential, lognormal, Maxwellian and Chi-squared with one degree
of freedom distributions. For a Rayleigh distribution, the
integral in equation (8) has the same value of the integral for a
exponential distribution. So, the mixed Gaussian-exponential and
the Gaussian-Rayleigh power spectrum are exactly the same. As we
can see in Table 1, the difference between various mixture
components is quite small for the mixed models, due to a very
small change in P$^{mix}$ for different components. For
\ensuremath{\alpha}$_{0}$\ensuremath{\sim}10$^{-3}$, the mixed
term, M(\ensuremath{\alpha}$_{0}$), is about 10$^{-3}$. To obtain
C$_{l}$ we have to multiply M(\ensuremath{\alpha}$_{o}$) by
k$^{n+1}$ (\ensuremath{\sim}10$^{-1}$)$^{n+1}$ and integrate in
dk. So, the difference in C$_{l}$ due to different mixture
components is about \ensuremath{\sim}10$^{-4}$-10$^{-5}$, too
small compared to the simulations errors
(\ensuremath{\sim}10$^{-3}$) for C$_{l}$ estimation in
$\ensuremath{\mu}K^2$. However, the difference between a mixed
non-Gaussian and a pure Gaussian field is evident in both
temperature and polarization CMBR power spectra. It is clearly
seen that the amplitude fluctuations change with the mixture
parameter \ensuremath{\alpha}$_{0}$, but the main ingredient of
the model seems to be the mixed (correlated) photon density field
considered in the last scattering surface, and not only the
statistic treatment of the phase correlation introduced in
P$^{mix}$. A practical way to discriminate between a pure Gaussian
(adiabatic) field and a mixed non-Gaussian one is comparing the
polarized component, the mean temperature fluctuations and the
peaks amplitude in the power spectrum.

\begin{table}
\centering
($\Delta$T/T)$_{rms}$ x10$^{-5}$\\
\begin{tabular}{cccccccc}
\tableline
\\
$\alpha_0$ & $G + Exp$ & $G + Ln$ & $G + Max$ & $G + \chi^2$\\
\\
\tableline
0 & 3.2941 & 3.2941 & 3.2942 & 3.2942\\
1.0E-05 & 3.2848 & 3.2847 & 3.2848 & 3.2848\\
5.0E-05 & 3.2478 & 3.2478 & 3.2480 & 3.2480\\
1.0E-04 & 3.2028 & 3.2027 & 3.2028 & 3.2028\\
5.0E-04 & 2.8850 & 2.8850 & 2.8850 & 2.8850\\
1.0E-03 & 2.6250 & 2.6247 & 2.6251 & 2.6251\\
1.3E-03 & 2.5603 & 2.5595 & 2.5605 & 2.5605\\
1.5E-03 & 2.5591 & 2.5579 & 2.5594 & 2.5593\\
1.7E-03 & 2.5915 & 2.5899 & 2.5919 & 2.5919\\
1.9E-03 & 2.6564 & 2.6543 & 2.6569 & 2.6568\\
2.1E-03 & 2.7517 & 2.7487 & 2.7519 & 2.7518\\
2.5E-03 & 3.0184 & 3.0195 & 3.0195 & 3.0193\\
5.0E-03 & 5.8409 & 5.8323 & 5.8433 & 5.8422\\
1.0E-02 & 12.6310 & 12.6413 & 12.6417 & 12.6371\\
6.0E-02 & 62.4921 & 62.7690 & 62.7797 & 62.6557\\
1.0E-01 & 79.8232 & 80.3811 & 80.4025 & 80.1531\\
\tableline
\end{tabular}
\caption{\small The mean temperature fluctuations, $(\Delta
T/T)_{rms} \times 10^{-5}$, estimated for some mixture components
normalised by COBE.}
\end{table}

Multicomponent models resulting in an excess of power in small
scales have already been investigated with the aid of CMBR
anisotropy simulations. For instance, Bucher et al. (2000) have
investigated two CDM isocurvature modes (with a neutrino
isocurvature and an isocurvature velocity perturbations),
evaluated by the linearized perturbation theory in distinct
regular and singular modes (rather than growing and decaing
modes), and have found a great variation of peaks intensity for a
different initial power spectra. Another approach of
multicomponent models is that of Gordon (2001) and Amendola et al
(2002).They consider a correlated field of adiabatic and
isocurvature perturbations produced during a period of
cosmological inflation, described in a generic power law spectrum,
and show that correlations can cause the acoustic peak height to
increase relative to the plateau of CMBR.

Our point is quite different of those from the above mentioned
authors. We show that it is possible to directly assess and
quantify the mixture of a correlated adiabatic and isocurvature
non-Gaussian field. Figure 5 shows the relative amplitude of the
most distinguished peaks (the first three peaks) for a
Gaussian-lognormal mixed model temperature power spectrum. This
plot clearly shows the difference in the relative amplitude of the
peaks for \ensuremath{\alpha}$_{0}$ \texttt{<} 3x10$^{-3}$. This
behaviour points to another possibility to extract information
from a CMBR power spectrum: the possibility of detecting weakly
mixed density fields, even if we can not exactly identify the
mixture components distribution.

\begin{figure}
\begin{center}
\plotone{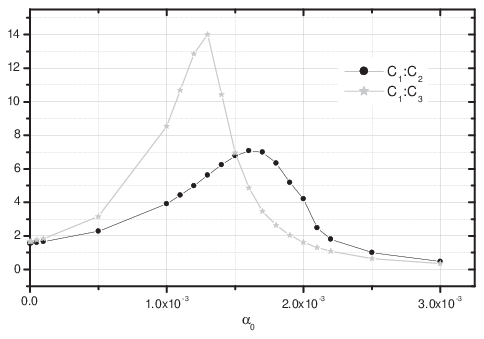} \caption{\small The relative amplitude of the
first three peaks for a Gaussian-lognormal mixed model temperature
power spectrum. $C_{i}:C_{j}$ means the \textit{ith} peak
amplitude divided by the \textit{jth} peak amplitude.}
\end{center}
\end{figure}

\section{Mixed Models and Cosmological Parameters}

A key question that will possibly be asked when trying to apply
this model to CMBR measurements is whether we can distinguish the
effects of a mixed field from those of variations in the
cosmological parameters. In order to answer this question and
quantify how a given cosmological parameter variation modify the
properties of an {\it a priori} unknown mixed field, we ran a number of realisations for a
wide range of cosmological parameters values, for both pure
Gaussian and mixed fields.

CMBR observational data are in good agreement with the basic
preferences of standard inflation scenarios: flat geometry
(\ensuremath{\Omega}$_{tot}$ \ensuremath{\sim} 1) and a nearly
scale invariant primeval spectrum (n \ensuremath{\sim} 1).
Nevertheless, degeneracies in parameter space prevent independent
determination of various cosmological parameters, like
\ensuremath{\Omega}$_{b}$, (the baryon density energy),
\ensuremath{\Omega}$_{CDM}$, (cold dark matter energy density),
\ensuremath{\Omega}$_{\ensuremath{\Lambda}}$ (vacuum energy
density) and H$_{0}$ (the Hubble constant) (Turner 1998;
Efstathiou 2001). Recent CMBR and large scale structure
observations suggest the Hubble constant H$_{0}$ to assume values
in the range (60 \texttt{<} H$_{0}$ \texttt{<}
70)(65\ensuremath{\pm}5) km/sec.Mpc (Turner 1998); and a positive
cosmological constant value in the range (0.065\texttt{<}
\ensuremath{\Omega}$_{\ensuremath{\Lambda}}$ \texttt{<}0.85)
(Efstathiou 2001; Spergel et al. 2003). The mass density of
baryons determined by Big Bang nucleossinthesys is
\ensuremath{\Omega}$_{b}$ =(0.019 \ensuremath{\pm} 0.01)h$^{-2}$,
which is in good agreement with CMBR observations (Stompor et al.
2001). To be consistent with these estimations, we ran another set
of realizations for a pure adiabatic Gaussian field, considering a
flat Universe and a range of possibilities for five cosmological
parameters: 0.8\texttt{<} n \texttt{<}1.2; 0.015
\texttt{<}\ensuremath{\Omega}$_{b}$\texttt{<} 0.03; 0.6
\texttt{<}\ensuremath{\Omega}$_{\ensuremath{\Lambda}}$\texttt{<}
0.8; and 60 \texttt{<}H$_{0}$\texttt{<} 80. The cold dark matter
density was set as 1-(\ensuremath{\Omega}$_{b}$ +
\ensuremath{\Omega}$_{\ensuremath{\Lambda}}$), ranging of 0.170
\texttt{<} \ensuremath{\Omega}$_{cdm}$\texttt{<} 0.385. The
temperature power spectra, simulated for a pure Gaussian field
generated with the above range of parameters are plotted in
Figures 6 and 7.

\begin{figure}
\begin{center}
\includegraphics[width=3.3in, height=1.8in]{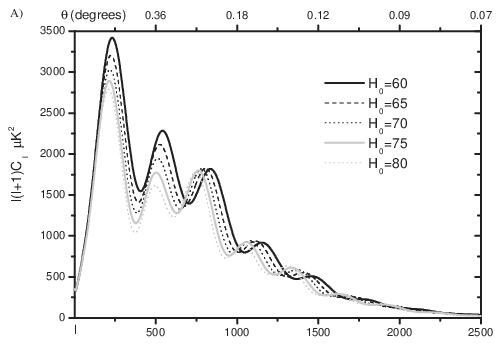}
\includegraphics[width=3.3in, height=1.8in]{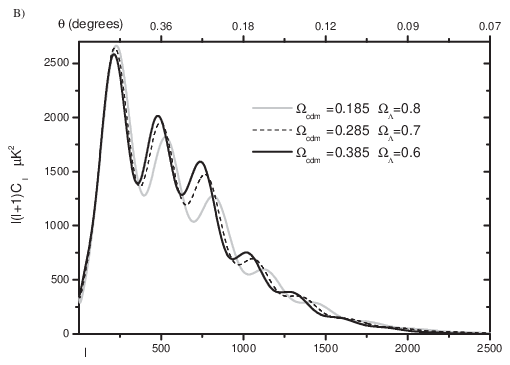}
\includegraphics[width=3.3in, height=1.8in]{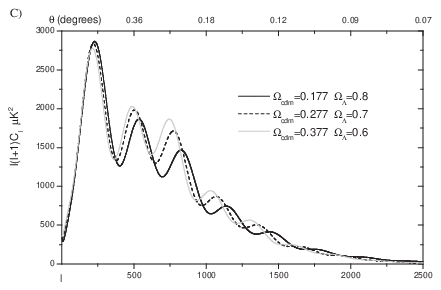}
\includegraphics[width=3.3in, height=1.8in]{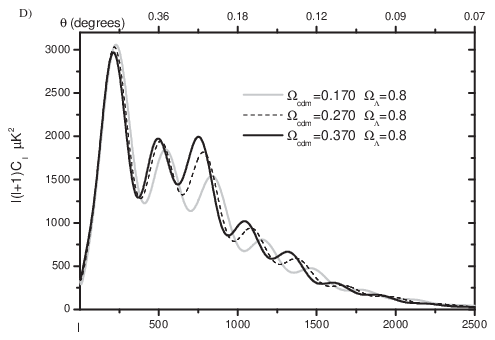}
\caption {\small The temperature power spectra simulated for a
pure Gaussian field combining a wide class of cosmological
parameters. In A, the curves show variations in the Hubble
constant for a model with $\Omega_b=0.03$, $\Omega_{cdm}=0.27$,
$\Omega_\Lambda=0.7$ and $n=1.1$. In B, C, and D, $\Omega_{b}$ is
set as 0.015, 0.023 and 0.03, respectively; $\Omega_{cdm}$ and
$\Omega_{\Lambda}$ is varying, while the spectral index is fixed
as 1.1, and $H_{0}=70$.}
\end{center}
\end{figure}

\begin{figure}
\begin{center}
\plotone{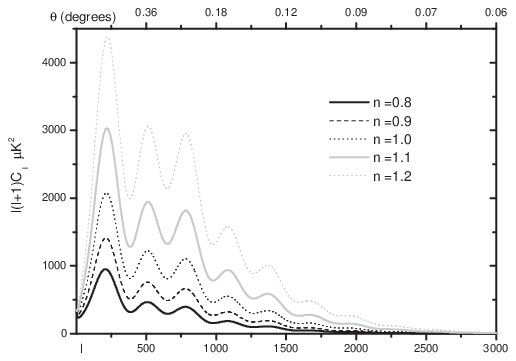} \caption {\small The temperature power spectra
simulated for a pure Gaussian field for a model with
$\Omega_{b}=0.03$, $\Omega_{cdm}=0.27$, $\Omega_\Lambda=0.7$,
$H_{0}=70$ and spectral index varying from 0.8 to 1.2.}
\end{center}
\end{figure}

We can see the effect of variations in the Hubble parameter in in
Figure 6A. Increasing H$_{0}$, the position of all peaks are
deviated towards smaller {\it l} while their amplitudes decrease.
For fixed \ensuremath{\Omega}$_{b}$, the effect of increasing
\ensuremath{\Omega}$_{cdm}$ while reducing
\ensuremath{\Omega}$_{\ensuremath{\Lambda}}$ is to shift the
second and third acoustic peaks positions to smaller {\it l} and
to higher amplitudes, while the primary peak features are barely
disturbed. The first peak has lower intensity for high
\ensuremath{\Omega}$_{cdm}$ (and low
\ensuremath{\Omega}$_{\ensuremath{\Lambda}}$), while higher order
peaks have higher intensity, as can be seen in Figures 6B, 6C and
6D. Figure 7 shows the effect of varying the spectral index, n, in
a pure adiabatic Gaussian field. As n increases, all the acoustic
peaks become intensified and slightly shifted to smaller {\it l}.

For all the combined cosmological parameters simulated we have
estimated the relative amplitude of the first three acoustic peaks
and the mean temperature fluctuation. The estimated values are
presented in Table 2. Comparing the values presented in Table 2
with the curves shown in Figure 6, we observe that the relative
amplitude of the peaks is lower than 2.1 for a wide combination of
cosmological parameters, while the relative amplitude is greater
than 2, for a mixed degree in the range: 5x10$^{-4}$\texttt{<}
\ensuremath{\alpha}$_{0}$ \texttt{<}2.1x10$^{-3}$. So, comparing
the peaks amplitude, a mixed (adiabatic and isocurvature) spectrum
is clearly distinct from a pure adiabatic one with a
different combination of cosmological parameters.

Figure 8 shows the behavior of temperature power spectrum for
mixed models according to the variations in the cosmological
parameters values, for five different cosmological models with
\ensuremath{\alpha}$_{0}$=1.5x10$^{-3}$. The behaviour of the
acoustic peaks is similar to that seen in Figure 1, with power
being transfered to higher {\it l}, while the super-degree scales
are not affected. Table 3 contains the relative amplitude of the
peaks and the mean temperature fluctuation for the seventeen
simulated mixed models, using
\ensuremath{\alpha}$_{0}$=5x10$^{-4}$.

\begin{table}
\centering
\begin{tabular}{cccc}
\tableline
\\
${\rm Model}$ & $C_1:C_2$ & $C_1:C_3$ & $(\Delta T/T)rms $\\
$ $ & $ $ & $ $ & $ \times10^{-5}$ \\
\tableline
$H_0$=70         $\Omega_b$=0.015\\
$\Omega_\Lambda$= 0.8       n= 1.1 &   1.46 &   2.09 & 3.057\\
\tableline
$H_0$=70         $\Omega_b$=0.015\\
$\Omega_\Lambda$= 0.7       n= 1.1 &   1.35 &   1.79 & 3.121\\
\tableline
$H_0$=70         $\Omega_b$=0.015\\
$\Omega_\Lambda$= 0.6       n= 1.1 &   1.28 &   1.62 & 3.148\\
\tableline
$H_0$=70         $\Omega_b$=0.023\\
$\Omega_\Lambda$= 0.8       n= 1.1 &   1.54 &   1.95 & 3.157\\
\tableline
$H_0$=70         $\Omega_b$=0.023\\
$\Omega_\Lambda$= 0.7       n= 1.1 &   1.43 &   1.66 & 3.221\\
\tableline
$H_0$=70         $\Omega_b$=0.023\\
$\Omega_\Lambda$= 0.6       n= 1.1 &   1.37 &   1.49 & 3.255\\
\tableline
$H_0$=70         $\Omega_b$=0.030\\
$\Omega_\Lambda$= 0.8       n= 1.1 &   1.65 &   1.98 & 3.226\\
\tableline
$H_0$=70         $\Omega_b$=0.030\\
$\Omega_\Lambda$= 0.7       n= 1.1 &   1.56 &   1.67 & 3.295\\
\tableline
$H_0$=70         $\Omega_b$=0.030\\
$\Omega_\Lambda$= 0.6       n= 1.1 &   1.50 &   1.49 & 3.326\\
\tableline
$H_0$=60         $\Omega_b$=0.030\\
$\Omega_\Lambda$= 0.7       n= 1.1 &   1.50 &   1.88 & 3.454\\
\tableline
$H_0$=65         $\Omega_b$=0.030\\
$\Omega_\Lambda$= 0.7       n= 1.1 &   1.51 &   1.76 & 3.368\\
\tableline
$H_0$=75         $\Omega_b$=0.030\\
$\Omega_\Lambda$= 0.7       n= 1.1 &   1.63 &   1.61 & 3.226\\
\tableline
$H_0$=80         $\Omega_b$=0.030\\
$\Omega_\Lambda$= 0.7       n= 1.1 &   1.73 &   1.57 & 3.170\\
\tableline
$H_0$=70         $\Omega_b$=0.030\\
$\Omega_\Lambda$= 0.7       n= 0.8 &   2.03 &   2.40 & 2.004\\
\tableline
$H_0$=70         $\Omega_b$=0.030\\
$\Omega_\Lambda$= 0.7       n= 0.9 &   1.86 &   2.12 & 2.349\\
\tableline
$H_0$=70         $\Omega_b$=0.030\\
$\Omega_\Lambda$= 0.7       n= 1.0 &   1.70 &   1.88 & 2.773\\
\tableline
$H_0$=70         $\Omega_b$=0.030\\
$\Omega_\Lambda$= 0.7       n= 1.2 &   1.43 &   1.48 & 3.929\\
\tableline
\end{tabular}
\caption{\small The relative amplitude of the peaks and the mean
temperature fluctuation estimated for a wide class of models for a
pure Gaussian field. }
\end{table}

\begin{figure}
\begin{center}
\plotone{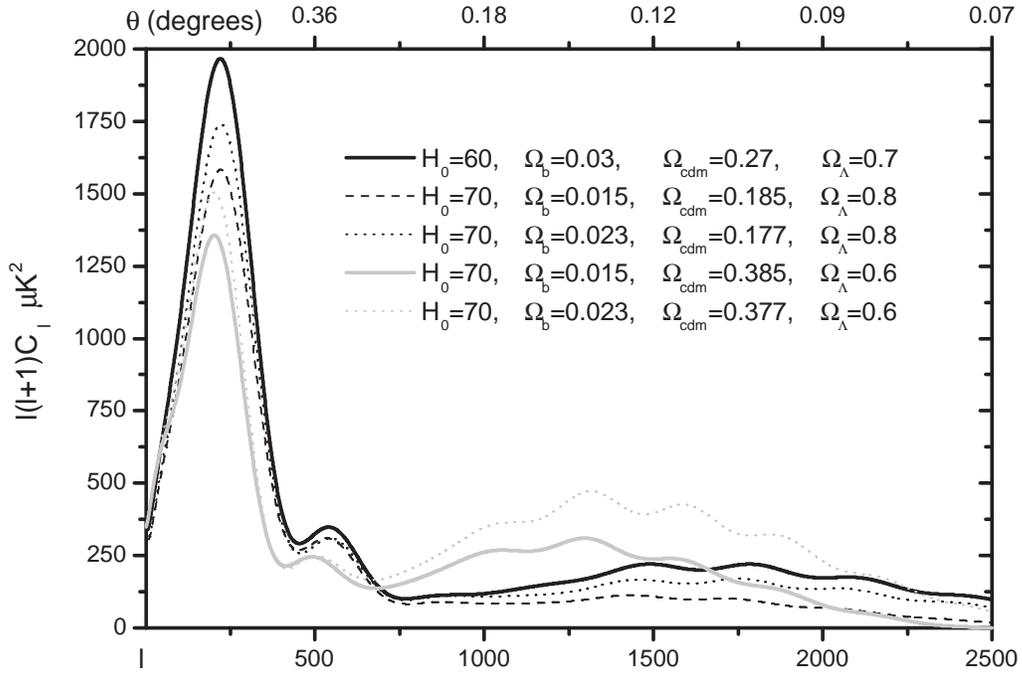} \caption {\small The temperature power spectra
simulated for a mixed Gaussian-lognormal field for some classes of
$\Omega_{b}$, $\Omega_{cdm}$, $\Omega_\Lambda$ and $H_0$. The
spectral index is set as 1.1, and the mixture degree is $1.5\times
10^{-3}$.}
\end{center}
\end{figure}

\begin{table}
\centering
\begin{tabular}{cccc}
\tableline
\\
${\rm Model}$ & $C_1:C_2$ & $C_1:C_3$ & $(\Delta T/T)rms $\\
$ $ & $ $ & $ $ & $ \times10^{-5}$ \\
\tableline
$H_0$=70         $\Omega_b$=0.015\\
$\Omega_\Lambda$= 0.8       n= 1.1 &   2.00 &   3.64 & 2.366\\
\tableline
$H_0$=70         $\Omega_b$=0.015\\
$\Omega_\Lambda$= 0.7       n= 1.1 &   1.91 &   3.33 & 2.386\\
\tableline
$H_0$=70         $\Omega_b$=0.015\\
$\Omega_\Lambda$= 0.6       n= 1.1 &   1.87 &   3.19 & 2.391\\
\tableline
$H_0$=70         $\Omega_b$=0.023\\
$\Omega_\Lambda$= 0.8       n= 1.1 &   2.14 &   3.44 & 2.449\\
\tableline
$H_0$=70         $\Omega_b$=0.023\\
$\Omega_\Lambda$= 0.7       n= 1.1 &   2.07 &   3.10 & 2.472\\
\tableline
$H_0$=70         $\Omega_b$=0.023\\
$\Omega_\Lambda$= 0.6       n= 1.1 &   2.05 &   2.95 & 2.503\\
\tableline
$H_0$=70         $\Omega_b$=0.030\\
$\Omega_\Lambda$= 0.8       n= 1.1 &   2.33 &   3.51 & 2.513\\
\tableline
$H_0$=70         $\Omega_b$=0.030\\
$\Omega_\Lambda$= 0.7       n= 1.1 &   2.29 &   3.15 & 2.557\\
\tableline
$H_0$=70         $\Omega_b$=0.030\\
$\Omega_\Lambda$= 0.6       n= 1.1 &   2.30 &   2.97 & 2.587\\
\tableline
$H_0$=60         $\Omega_b$=0.030\\
$\Omega_\Lambda$= 0.7       n= 1.1 &   2.07 &   3.32 & 2.604\\
\tableline
$H_0$=65         $\Omega_b$=0.030\\
$\Omega_\Lambda$= 0.7       n= 1.1 &   2.16 &   3.20 & 2.565\\
\tableline
$H_0$=75         $\Omega_b$=0.030\\
$\Omega_\Lambda$= 0.7       n= 1.1 &   2.49 &   3.14 & 2.547\\
\tableline
$H_0$=80         $\Omega_b$=0.030\\
$\Omega_\Lambda$= 0.7       n= 1.1 &   2.76 &   3.18 & 2.565\\
\tableline
$H_0$=70         $\Omega_b$=0.030\\
$\Omega_\Lambda$= 0.7       n= 0.8 &   3.00 &   4.55 & 1.695\\
\tableline
$H_0$=70         $\Omega_b$=0.030\\
$\Omega_\Lambda$= 0.7       n= 0.9 &   2.74 &   4.02 & 1.930\\
\tableline
$H_0$=70         $\Omega_b$=0.030\\
$\Omega_\Lambda$= 0.7       n= 1.0 &   2.50 &   3.55 & 4.664\\
\tableline
$H_0$=70         $\Omega_b$=0.030\\
$\Omega_\Lambda$= 0.7       n= 1.2 &   2.11 &   2.79 & 2.976\\
\tableline
\end{tabular}
\caption{\small The relative amplitude of the peaks and the mean
temperature fluctuation estimated for a wide class of models for a
mixed Gaussian-lognormal field with a mixed degree of $5\times
10^{-4}$.}
\end{table}

Comparing the values in Tables 2 and 3, we can verify the
difference between the relative amplitude for mixed and pure
models, for a quite large combination of cosmological parameters.
It seems clear that the effect of mixed models upon the acoustic
peaks amplitude is more pronounced than those achieved through
variations of the cosmological parameters in a pure model. Besides
that, once the above cosmological parameters are determined with
better precision, power spectrum examination can be used to
identify a mixed density field and estimate the mixture degree by
comparing the peaks intensity.

In order to quantify a possible non-gaussian contribution to the
fluctuations field, we have made a maximum likelihood approach to
power spectrum estimated by several classes of CMB experiments.
Our best estimate of the angular power spectrum for the CMB is
shown in Figure 9 for a combination of various CMB data prior to
WMAP, and, in Figure 10, the best standard gaussian and the bests
estimated mixed model fits for the first-year WMAP data. The
$\chi^2$ test shows that the contribution of an isocurvature
fluctuation field is not ruled out in actual CMB observations.

\begin{figure}
\begin{center}
\plotone{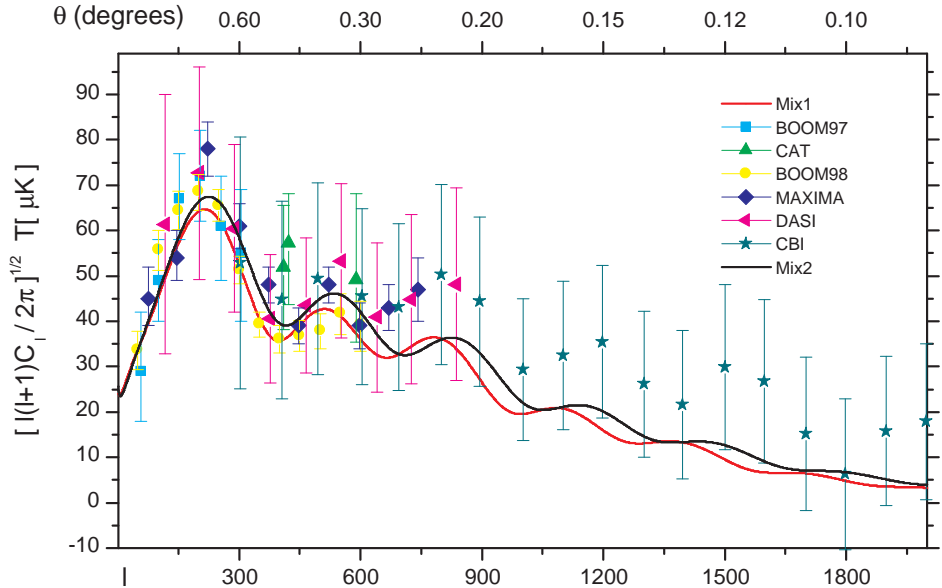} \caption {\small Temperature fluctuation
estimated for some classes of CMBR experiments and the mixed model
estimative for \ensuremath{\alpha}$_{0}$\texttt{=}5x10$^{-4}$ in
two combinations of cosmological parameters, Mix1:
$\Omega_b=0.023$, $\Omega_{cdm}=0.177$, $\Omega_\Lambda=0.8$;
$n=1.1$ and $H_0=70 km/sMpc $; and Mix2: $\Omega_b=0.03$,
$\Omega_{cdm}=0.27$, $\Omega_\Lambda=0.7$; $n=1.1$ and $H_0=70
km/sMpc $. The reduced $\chi^2$ estimated are 1.010 and 1.000,
respectively.}
\end{center}
\end{figure}

\begin{figure}
\begin{center}
\plotone{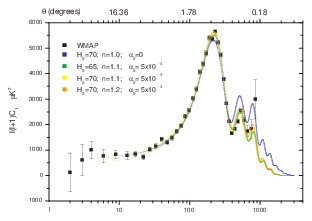} \caption {\small The angular power spectrum
estimated for the WMAP mission, the predictions of the standard
$\Lambda-CDM$ model and tree best-fit $\Lambda-CDM$ mixed model
for the CMBR. In all models illustrated, the best cosmological
parameters combination found was: $\Omega_b=0.03$,
$\Omega_{cdm}=0.27$, $\Omega_\Lambda=0.7$. The reduced $\chi^2$
estimated are 13.399 for $H_0=70 km/sMpc$; $n=1.0$ and
$\alpha_0=0$; 1.723 for $H_0=65 km/sMpc$; $n=1.1$ and
$\alpha_0=5\times10^{-4}$; 2.260 for $H_0=70 km/sMpc$; $n=1.2$ and
$\alpha_0=5\times10^{-4}$ and 2.546 for $H_0=70 km/sMpc$; $n=1.1$
and $\alpha_0=5\times10^{-4}$.}
\end{center}
\end{figure}

\section{Discussion}
Assuming the hypothesis that the initial fluctuation field is the
result of weakly correlated adiabatic and isocurvature mixed
fields, we made a number of realisations of CMBR temperature and
polarisation power spectra and estimated the mean temperature
fluctuations combining Gaussian, Exponential, Lognormal, Rayleigh,
Maxwellian and a Chi-squared distributions. The contribution of a
second field is estimated by the distortion of the power spectrum
and is also considered a correlated amplitude for the mixed field.
Some important results were obtained. We show that it is possible
to directly assess and quantify the mixture of a correlated
adiabatic and isocurvature non-Gaussian field. The simulations
clearly show the difference in the relative amplitude of the
acoustic peaks for a mixed correlated model. This behaviour points
to another possibility to extract information from a CMBR power
spectrum: the possibility of detecting weakly mixed density
fields, even if we can not exactly identify the mixture components
distribution. The simulations show that the influence of the
specific statistics of the second component in the mixed field is
not so important as the cross-correlation between the amplitudes
of both fields. This seems to be very important in the CMBR power
spectrum estimation. We claim that a physical mechanism
responsible for generation of both field could result in a
distinctive signature in CMBR. We also show the results are not
strongly affected by the choice of the cosmological parameters and
hence the characteristic behaviour of the acoustic peaks amplitude
and the polarized component can be used as a cosmological test for
the nature of the primordial density field. Indeed, the
predictions of mixture models are very distinct from the pure
density fields, especially for small angular scales.

Also, recent temperature fluctuation CMBR observational data do
not rule out a mixture component in a small level of
non-Gaussianity, with a mixture coefficient of $\sim \times
10^{-4}$, as can be seen in Figure 9 and Figure 10. With the new
generation of the CMBR experiments, especially the expected
satellite mission Planck (Tauber 2000), we expect to be able to
compare more observations in small scales, with better signal to
noise and larger sky coverage, to the predictions of our class of
mixed models and estimate the physical mechanisms responsible for
structure formation. In despite of the degeneration of the power
spectrum for mixed PDF, we expect to better estimate the
statistical description of fluctuations in a mixed scenario by
carrying out the investigation, in a non-gaussian context, of the
average correlation function and the correlation function for high
amplitude peaks (Andrade et al., 2003)

\acknowledgments{A.P.A.A acknowledges the fellowship from CAPES
(Coordena\c {c}\~ao de Aperfei\c {c}oamento de Pessoal de Ensino
Superior). C.A.W. acknowledges a research fellowship received by
Conselho Nacional de Desenvolvimento Cient\'ifico e Tecnol\'ogico
(CNPq) for a research grant 300409/97-4. The authors acknowledge
Edmund Bertschinger for the use of the COSMICS package, funded by
NSF under grant AST-9318185.}

%========================================================================

%========================================================================

\end{document}